\documentclass[sigconf, nonacm]{acmart}

\usepackage{graphicx}
\usepackage{xcolor}
\usepackage{booktabs}
\usepackage{listings}
\usepackage{tikz}
\usepackage{amsmath}
\usepackage{enumitem}
\usepackage{float}
\usepackage{tabularx}
\usepackage{microtype}

\usetikzlibrary{arrows.meta, positioning, calc, fit, backgrounds}

\definecolor{aegonblue}{RGB}{41, 98, 168}
\definecolor{aegonlight}{RGB}{230, 240, 255}
\definecolor{codebg}{RGB}{246, 247, 249}

\tikzset{
  anode/.style={rectangle, rounded corners=3pt, draw=aegonblue, fill=aegonlight, font=\small, minimum width=3.6cm, minimum height=0.7cm, align=center, text=black},
  anodebold/.style={anode, font=\small\bfseries},
  aarrow/.style={-{Stealth[length=5pt]}, aegonblue, thick},
  adasharrow/.style={-{Stealth[length=5pt]}, aegonblue!70, thick, dashed},
  adotarrow/.style={-{Stealth[length=4pt]}, aegonblue!60, densely dotted, thin},
  alabel/.style={font=\scriptsize, fill=white, inner sep=2pt, text=black},
}

\setlist[enumerate]{topsep=4pt, itemsep=3pt, parsep=0pt}
\setlist[itemize]{topsep=4pt, itemsep=3pt, parsep=0pt}

\begin{document}


\title{Aegon: Auditable AI Content Access with Ledger-Bound Tokens and Hardware-Attested Mobile Receipts}

\author{Amrish Baskaran \and Nirbhay Pherwani \and Raghul Krishnan}

\begin{abstract}
Recent standards such as RSL~\cite{rsl} address AI content \emph{policy declaration}---telling AI systems what the licensing terms are. However, no existing system provides \emph{audit infrastructure}---tamper-evident licensing transaction records with independently verifiable proofs that those records have not been retroactively modified. We describe Aegon, a protocol that extends standard JWT tokens with content-specific licensing claims and maintains a Certificate Transparency-style~\cite{rfc6962} Merkle tree over an append-only transaction ledger, enabling third-party auditors to independently verify that specific content licensing transactions were recorded and have not been retroactively modified. Publishers validate tokens at the edge using standard JWKS with no broker dependency in the content delivery path. A signed provenance event log tracks content through AI transformation stages (chunking, embedding, retrieval, citation), bound to ledger entries by transaction ID. We further describe hardware-attested compliance receipts for on-device Android AI agents using StrongBox secure element attestation~\cite{strongbox}---to our knowledge, the first application of hardware-attested compliance receipts to AI content licensing. Existing DRM systems use hardware-backed keys for content decryption but do not produce verifiable compliance receipts for audit trails. We describe a reference architecture and an evaluation methodology for measuring protocol overhead. The protocol runs entirely over standard HTTPS and is designed to complement existing licensing standards rather than replace them.
\end{abstract}

\keywords{AI content licensing, audit ledger, Certificate Transparency, hardware attestation, compliance protocol, ledger-bound tokens}

\renewcommand{\shortauthors}{Baskaran, Pherwani, and Krishnan}

\maketitle

\section{Introduction}
\label{sec:intro}

\subsection{Motivation}

AI systems---large language models used for training, inference, and retrieval-augmented generation (RAG)---depend heavily on web content. Yet no standard mechanism exists for AI platforms to access content with verifiable licensing, and no standard mechanism exists for publishers to enforce or monetize access rights.

The result is a coordination failure:
\begin{itemize}
  \item Publishers block all AI traffic or allow unchecked access with no compensation
  \item AI platforms scrape without licenses, facing ongoing lawsuits with billions in potential exposure~\cite{nyt-v-openai, getty-v-stability, authorsguild-v-openai}
  \item Bespoke licensing negotiations take months and do not scale~\cite{newscorp-openai}
  \item No audit trail exists to prove content was accessed legally
\end{itemize}

The same problem exists at a second layer: on-device AI agents (Gemini Nano~\cite{gemini-nano}, Llama~\cite{llama3}) access web content during inference on mobile devices with no licensing infrastructure whatsoever.

\subsection{Contributions}

Aegon builds on standard JWT~\cite{rfc7519} and JWKS~\cite{rfc7517} infrastructure and makes contributions in two areas that, to our knowledge, have no prior art in the content licensing domain:

\textbf{Contribution~1---Auditable Web Protocol Layer:}
\begin{enumerate}
  \item Content-specific licensing claims extending standard JWT---the claim semantics are the contribution, not the validation mechanism.
  \item A Certificate Transparency-style~\cite{rfc6962} Merkle tree over an append-only transaction ledger, enabling third-party audit verification via Signed Tree Heads and inclusion proofs---to our knowledge, the first application of the CT audit model to content licensing.
  \item A signed provenance event log tracking content through AI-specific transformation stages, bound to ledger entries by transaction ID.
\end{enumerate}

\textbf{Contribution~2---Android Attestation Layer:}
\begin{enumerate}
  \item Hardware-attested compliance receipts for on-device AI agents using Android StrongBox secure element~\cite{strongbox}---to our knowledge, the first application of hardware-attested compliance receipts to AI content licensing on mobile devices.
  \item Offline proof batching with replay resistance for intermittent mobile connectivity.
  \item Broker-side verification of Key Attestation~\cite{key-attestation} certificate chains rooted at Google hardware attestation CA.
\end{enumerate}

Existing licensing standards (RSL~\cite{rsl}, AIBDP~\cite{aibdp}) address policy declaration. Content access protocols (Peek-Then-Pay~\cite{peek-then-pay}, PEAC~\cite{peac}) address the licensing transaction flow. Aegon's distinct contribution is the \emph{audit infrastructure}---the Merkle-committed ledger, the cryptographically signed provenance chain, and the hardware-attested mobile receipts---that enables independent verification of licensing transactions and tamper-evident provenance records after the fact.

\subsection{Analogous Infrastructure}

Aegon's position in the AI content stack is analogous to prior HTTP-based coordination infrastructure: OAuth~2.0~\cite{rfc6749} standardized delegated authentication; Certificate Transparency~\cite{rfc6962} introduced Merkle tree audit logs for TLS certificates; PCI~DSS established a compliance standard that companies adopt because they need to demonstrate compliance. Aegon adopts CT's Merkle tree audit construction and Signed Tree Head pattern as a CT-inspired audit design; it does not claim to replicate the full CT ecosystem, and the broker remains a trusted party rather than a trustless log.

\section{Background and Related Work}
\label{sec:related}

The AI content licensing space has seen rapid development since 2024. We organize related work into six categories.

\subsection{Emerging Licensing Standards}

\textbf{RSL~1.0}~\cite{rsl}: Published December 2025 by the co-creators of RSS, RSL defines a machine-readable XML vocabulary for content licensing, a Crawler Authorization Protocol (CAP) using license tokens in HTTP headers, and an Open License Protocol (OLP) extending OAuth~2.0. Endorsed by 1,500+ organizations including Cloudflare and Akamai. RSL addresses \emph{policy declaration}; Aegon addresses \emph{auditability of licensing transactions and tamper-evident provenance evidence}. RSL has no audit ledger, no Merkle proofs, no provenance tracking, and no hardware attestation.

\textbf{AIBDP}~\cite{aibdp}: An IETF Internet-Draft (August 2025) defining machine-readable AI permissions via HTML meta tags, DNS TXT records, and HTTP headers. A declaration standard with no enforcement mechanism.

\textbf{Agentic JWT}~\cite{agentic-jwt}: An IETF Internet-Draft (December 2025) extending OAuth~2.0 for AI agents with cryptographic agent identity. Addresses \emph{who the agent is}; Aegon addresses \emph{what content the agent can access}. Complementary.

\subsection{Content Access Protocols}

\textbf{Peek-Then-Pay}~\cite{peek-then-pay}: An HTTP 203-based protocol with JWT licenses and edge validation. The most architecturally similar system to Aegon. Key difference: Peek-Then-Pay optimizes the \emph{discovery and pricing} flow; Aegon provides the \emph{audit and licensing evidence} infrastructure.

\textbf{PEAC}~\cite{peac}: A verifiable receipt system with compliance proofs and payment integration. PEAC's receipts overlap with Aegon's audit trail, but PEAC lacks Merkle tree verifiability and hardware attestation.

\textbf{OpenAttribution}~\cite{openattribution}: An open-source framework with DID-based agent identity and session-based content event tracking through AI pipeline stages. Conceptually similar to Aegon's provenance log, but without cryptographic binding to Merkle-committed ledger entries.

\textbf{TollBit}~\cite{tollbit}: A commercial bot paywall and licensing platform (\$30.9M total funding including \$24M Series A, 500+ publisher sites). Closed-source SaaS without a public protocol specification or audit ledger.

\textbf{Cashmere}~\cite{cashmere}: A commercial data infrastructure platform (\$5M seed, January 2026) with live deployments at Wiley and Harvard Business Publishing, and Perplexity as both investor and customer. Provides token-level access control, rights management, usage tracking, and revenue settlement. Closed-source with no open protocol specification, no Merkle audit ledger, and no hardware attestation. Its commercial traction validates demand for the infrastructure Aegon describes.

\subsection{Traditional Access Control and DRM}

\textbf{robots.txt}~\cite{rfc9309}: Crawler exclusion with no authentication or licensing.

\textbf{OAuth~2.0}~\cite{rfc6749, oidc}: The standard for delegated authentication. Aegon inherits JWKS-based validation~\cite{rfc7517}. Aegon's contribution is the content licensing claim semantics and audit ledger binding.

Traditional DRM (Widevine~\cite{widevine}, FairPlay, PlayReady) protects media at the delivery layer. Aegon differs: (1)~no content encryption---audit trails, not access prevention; (2)~AI data pipelines, not media playback; (3)~JWKS verification, not license server roundtrip.

\subsection{Content Provenance}

C2PA~\cite{c2pa} defines cryptographic provenance for digital assets. C2PA v2.x extended coverage to include text and AI pipeline assertions; Aegon complements rather than replaces it. The key distinction: C2PA is \emph{creator-attested at authorship time} (file-level); Aegon is \emph{platform-attested at consumption time}, bound to a licensing transaction (pipeline-level). VeriTrail~\cite{veritrail} (ICLR 2026) traces provenance for hallucination detection, not licensing. FG-Trac~\cite{fg-trac} provides sample-level ML training traceability with cryptographic commitments, but at the training level. Information Isotopes~\cite{info-isotopes} enable post-hoc detection of training data usage---complementary to Aegon's ex-ante licensing.

\subsection{Blockchain-Based Approaches}

IBIS~\cite{ibis} uses blockchain for AI training copyright compliance. Content ARCs~\cite{content-arcs} combine C2PA, ODRL, and DLT. Both use blockchain consensus with higher overhead~\cite{blockchain-drm, timestamping}. Aegon uses a centralized append-only ledger with a CT-style~\cite{rfc6962} Merkle tree---the same verifiability without consensus overhead, at database write speeds. CT proved this model works at internet scale.

\subsection{Mobile Attestation}

\textbf{Android Key Attestation}~\cite{key-attestation}: Verifies hardware-backed key generation (StrongBox~\cite{strongbox} or KeyMint~\cite{keymint}). Used in Play Integrity API~\cite{play-integrity}. Not previously applied to content licensing. \textbf{Apple App Attest}~\cite{app-attest}: Apple's equivalent; future work. No prior work applies hardware attestation to AI content licensing.

\section{Threat Model}
\label{sec:threat}

\subsection{Trust Assumptions}

\textbf{The broker is a trusted third party.} It holds the signing key, maintains the ledger, and issues tokens. If the broker is compromised, token integrity, ledger integrity, and billing accuracy are all at risk. Mitigations: HSM-backed signing keys (AWS KMS), append-only ledger with write-once rows, and third-party audit access. This trust requirement is equivalent to trusting a Certificate Authority in the CT model or an OAuth~2.0 authorization server---both accepted as reasonable trust anchors for internet-scale infrastructure. A fully decentralized design (eliminating broker trust) is out of scope for this paper but discussed in Section~\ref{sec:future}.

\textbf{Publishers and platforms are mutually distrusting.} Publishers want proof that platforms respect license terms. Platforms want proof that publishers serve the content they paid for.

\subsection{Web Layer Adversaries}

\begin{itemize}
  \item \emph{Unlicensed platform:} Accesses content without a valid token---detected by publisher-side JWT validation
  \item \emph{Token replayer:} Reuses a token for different content or beyond expiry---detected by \texttt{jti} deduplication (Section~\ref{sec:replay}) and \texttt{exp} claim checking
  \item \emph{Caching violator:} Caches beyond licensed TTL---detectable via provenance event logs; enforcement is contractual in v1
  \item \emph{Training violator:} Uses content for training when \texttt{training\_\allowbreak{}allowed = false}---detectable only if the platform reports honestly; cryptographic enforcement is future work
  \item \emph{Parallel pipeline:} A platform could fetch content via Aegon (clean provenance chain recorded) while simultaneously copying raw text into a separate training pipeline outside the SDK---detectable only via contractual audit rights and statistical anomaly detection. This limitation is analogous to PCI~DSS, where a determined adversary can process card data outside the compliant system; PCI~DSS addresses this through contractual audit rights, anomaly detection, and penalties rather than cryptographic prevention of all side-channels. V1 takes the same approach; ZK and TEE-based enforcement are future work (Section~\ref{sec:future}).
  \item \emph{Timestamp manipulator:} Provenance event timestamps are platform-generated and could be backdated. Mitigation: the broker records a server-side receipt timestamp on arrival; gaps beyond a threshold are flagged
\end{itemize}

\subsection{Web Layer Security Goals}

\begin{itemize}
  \item \textbf{Token integrity:} Modification of any JWT claim is detectable via JWKS signature verification
  \item \textbf{Offline validation:} Publishers verify tokens using cached JWKS without contacting the broker per-request
  \item \textbf{Post-facto auditability:} Auditors verify transaction existence via Merkle inclusion proofs against a published Signed Tree Head---analogous to CT monitors~\cite{rfc6962}
  \item \textbf{Non-repudiation:} Once a token is issued and recorded, the broker cannot deny issuance
\end{itemize}

\subsection{Android Layer Adversaries}

\begin{itemize}
  \item \emph{Rooted/unlocked device:} Key Attestation includes \linebreak \mbox{\texttt{verifiedBootState}}; broker rejects receipts from unlocked bootloaders
  \item \emph{Receipt replayer:} Broker deduplicates on \texttt{receipt\_id}
  \item \emph{Key extraction:} StrongBox provides hardware key isolation; keys are non-exportable
  \item \emph{Modified app:} Key Attestation proves key hardware-binding but \textbf{does not prove app integrity}. Combining with Play Integrity API~\cite{play-integrity} is future work.
\end{itemize}

\subsection{Android Layer Security Goals}

\begin{itemize}
  \item \textbf{Hardware binding:} Device keys generated inside StrongBox secure element cannot be exported
  \item \textbf{Attestation chain validity:} Broker verifies the full certificate chain back to Google hardware attestation root CA
  \item \textbf{Replay resistance:} Unique \texttt{receipt\_id} per receipt; broker deduplicates
  \item \textbf{Offline resilience:} Receipts queued locally when offline; \texttt{timestamp} field enables detection of excessive submission delay
\end{itemize}

\subsection{What This Protocol Does Not Guarantee}

\textbf{Content DRM:} Aegon does not encrypt content. \textbf{Training prevention:} A determined adversary can use content for training and omit provenance events. \textbf{Provenance completeness:} The chain is tamper-evident for logged events but cannot guarantee completeness against an adversarial SDK operator.

\section{System Design}
\label{sec:design}

\subsection{Actors and Architecture}

\textbf{AI Platform:} Requests licenses before accessing content. Integrates via a traditional platform SDK (Python, Node.js) or via an MCP~\cite{mcp} server that exposes licensing as native tool calls---any MCP-compatible AI agent acquires licenses without protocol-specific integration beyond standard MCP tool wiring.

\textbf{Publisher:} Validates tokens at the edge via JWKS using copy-paste Cloudflare Workers middleware---no SDK installation, no broker dependency in the content delivery path.

\textbf{Aegon Broker:} Issues signed JWT licenses. Maintains the append-only transaction ledger and Merkle tree. Provides JWKS endpoint. Publishes Signed Tree Heads.

\textbf{Mobile AI Agent (Android):} On-device LLM requesting licensed content. Generates hardware-attested compliance receipts signed by StrongBox~\cite{strongbox}, with offline-capable receipt batching via encrypted local storage, per-publisher pseudonymous identifiers to prevent cross-publisher device tracking, and graceful KeyMint~\cite{keymint} fallback on devices without StrongBox.

\textbf{Integration model.} The protocol is designed for near-zero adoption friction. Platforms connect via standard MCP tool calls or a lightweight SDK, without any protocol-specific integration; publishers deploy a single Cloudflare Workers script that caches JWKS and validates tokens at the CDN edge. Neither side requires a broker roundtrip on the content delivery path.

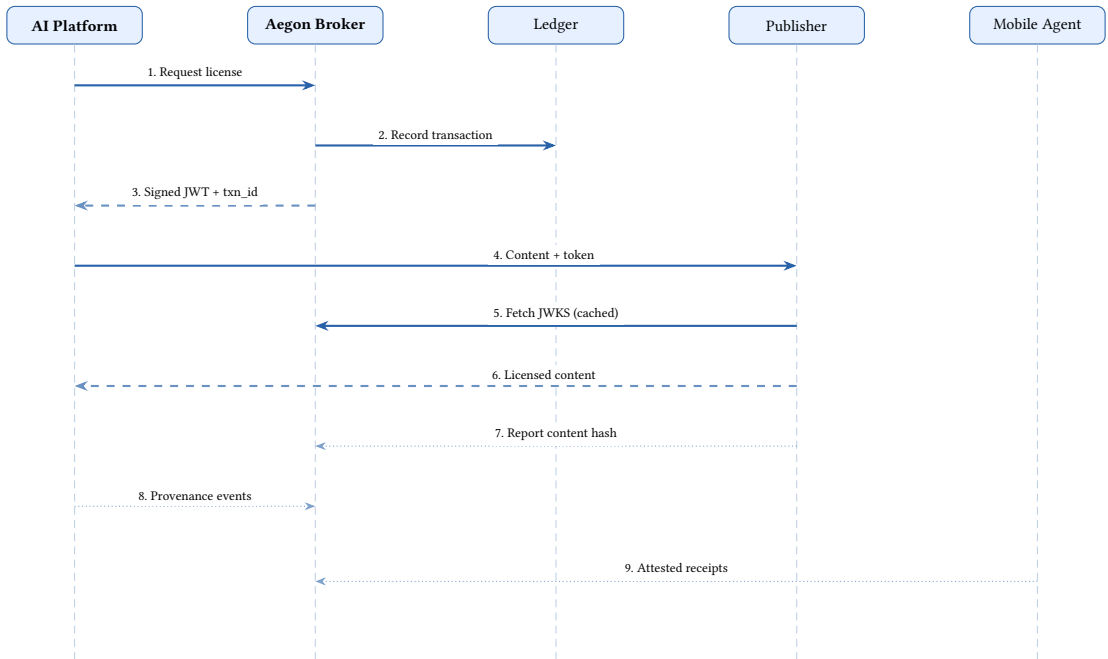
\begin{figure*}[t]
\centering
\begin{tikzpicture}
  \node[anodebold, minimum width=1.8cm, minimum height=0.5cm, font=\scriptsize\bfseries] (P) at (0,0) {AI Platform};
  \node[anodebold, minimum width=1.8cm, minimum height=0.5cm, font=\scriptsize\bfseries] (B) at (3.2,0) {Aegon Broker};
  \node[anode, minimum width=1.8cm, minimum height=0.5cm, font=\scriptsize] (L) at (6.4,0) {Ledger};
  \node[anode, minimum width=1.8cm, minimum height=0.5cm, font=\scriptsize] (Pub) at (9.6,0) {Publisher};
  \node[anode, minimum width=1.8cm, minimum height=0.5cm, font=\scriptsize] (M) at (12.8,0) {Mobile Agent};

  \draw[densely dashed, aegonblue!30, thin] (P.south) -- ++(0,-8.2);
  \draw[densely dashed, aegonblue!30, thin] (B.south) -- ++(0,-8.2);
  \draw[densely dashed, aegonblue!30, thin] (L.south) -- ++(0,-8.2);
  \draw[densely dashed, aegonblue!30, thin] (Pub.south) -- ++(0,-8.2);
  \draw[densely dashed, aegonblue!30, thin] (M.south) -- ++(0,-8.2);

  \draw[aarrow] (0,-0.8) -- node[alabel, above, font=\tiny] {1.~Request license} (3.2,-0.8);
  \draw[aarrow] (3.2,-1.6) -- node[alabel, above, font=\tiny] {2.~Record transaction} (6.4,-1.6);
  \draw[adasharrow] (3.2,-2.4) -- node[alabel, above, font=\tiny] {3.~Signed JWT + txn\_id} (0,-2.4);
  \draw[aarrow] (0,-3.2) -- node[alabel, above, font=\tiny, pos=0.65] {4.~Content + token} (9.6,-3.2);
  \draw[aarrow] (9.6,-4.0) -- node[alabel, above, font=\tiny] {5.~Fetch JWKS (cached)} (3.2,-4.0);
  \draw[adasharrow] (9.6,-4.8) -- node[alabel, above, font=\tiny, pos=0.35] {6.~Licensed content} (0,-4.8);
  \draw[adotarrow] (9.6,-5.6) -- node[alabel, above, font=\tiny] {7.~Report content hash} (3.2,-5.6);
  \draw[adotarrow] (0,-6.4) -- node[alabel, above, font=\tiny] {8.~Provenance events} (3.2,-6.4);
  \draw[adotarrow] (12.8,-7.4) -- node[alabel, above, font=\tiny] {9.~Attested receipts} (3.2,-7.4);
\end{tikzpicture}
\caption{System architecture showing the four actors and numbered data flows. Solid arrows are synchronous requests; dashed arrows are responses; dotted arrows are asynchronous reporting. All communication is standard HTTPS.}
\label{fig:architecture}
\end{figure*}

\bigskip

\subsection{Licensing Model}

Licenses are parameterized along three dimensions (Table~\ref{tab:licensing}). Scope defines what content the platform may access. License type controls temporal constraints on that access. Compliance flags are cryptographically bound to the token and visible to auditors; enforcement against violations is contractual in v1 (see Section~\ref{sec:future}).

\begin{table}[htb]
\centering
\begin{tabularx}{\columnwidth}{l>{\raggedright\arraybackslash}X}
\toprule
\textbf{Dimension} & \textbf{Values} \\
\midrule
Scopes & \texttt{metadata\_only}, \texttt{excerpt}, \texttt{full\_article\_html}, \texttt{structured\_json}, \texttt{training\_corpus} \\
License types & \texttt{single\_use}, \texttt{session}, \texttt{time\_bound\_cache}, \texttt{training\_corpus} \\
Compliance flags & \texttt{training\_allowed}, \texttt{attribution\_required}, \texttt{provenance\_required} \\
\bottomrule
\end{tabularx}
\caption{Licensing model dimensions.}
\label{tab:licensing}
\end{table}

\subsection{Token Format}

Aegon license tokens are standard JWTs~\cite{rfc7519} with content-specific claims (Listing~\ref{lst:token}):

\begin{lstlisting}[language={},float=htb,caption={JWT license token format.},label={lst:token}]
{
  "iss": "broker.aegon.ai",
  "sub": "platform_id",
  "aud": "publisher.com",
  "exp": 1798761600,
  "jti": "txn_abc123",
  "aegon_version": "1.0",
  "aegon_resource_url": "https://publisher.com/article/123",
  "aegon_scope": "full_article_html",
  "aegon_license_type": "session",
  "aegon_training_allowed": false,
  "aegon_attribution_required": true
}
\end{lstlisting}

\textbf{Design decision---content hash is NOT in the token.} The broker issues the token before the platform fetches content. The content hash cannot be known at issuance time. Instead, the publisher computes the hash post-delivery and records it in the ledger as a separate entry linked to the \texttt{jti}. This enables post-facto verification.

\textbf{What Aegon adds to standard JWT:} The \texttt{aegon\_*} claims carry licensing semantics. The validation mechanism (JWKS) is standard; the claim semantics and the system around them are the contribution.

\section{Web Protocol}
\label{sec:protocol}

\subsection{License Request Flow}

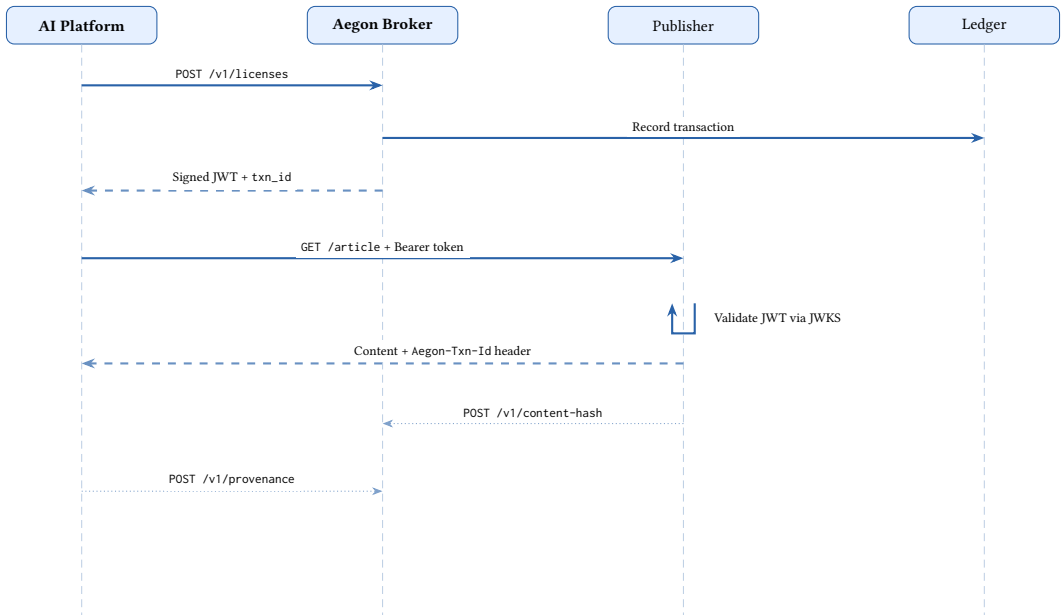
\begin{figure*}[t]
\centering
\begin{tikzpicture}
  \node[anodebold, minimum width=2cm, minimum height=0.5cm, font=\scriptsize\bfseries] (P) at (0,0) {AI Platform};
  \node[anodebold, minimum width=2cm, minimum height=0.5cm, font=\scriptsize\bfseries] (B) at (4,0) {Aegon Broker};
  \node[anode, minimum width=2cm, minimum height=0.5cm, font=\scriptsize] (Pub) at (8,0) {Publisher};
  \node[anode, minimum width=2cm, minimum height=0.5cm, font=\scriptsize] (L) at (12,0) {Ledger};

  \draw[densely dashed, aegonblue!30, thin] (P.south) -- ++(0,-7.6);
  \draw[densely dashed, aegonblue!30, thin] (B.south) -- ++(0,-7.6);
  \draw[densely dashed, aegonblue!30, thin] (Pub.south) -- ++(0,-7.6);
  \draw[densely dashed, aegonblue!30, thin] (L.south) -- ++(0,-7.6);

  \draw[aarrow] (0,-0.8) -- node[alabel, above, font=\tiny] {\texttt{POST /v1/licenses}} (4,-0.8);
  \draw[aarrow] (4,-1.5) -- node[alabel, above, font=\tiny] {Record transaction} (12,-1.5);
  \draw[adasharrow] (4,-2.2) -- node[alabel, above, font=\tiny] {Signed JWT + \texttt{txn\_id}} (0,-2.2);

  \draw[aarrow] (0,-3.1) -- node[alabel, above, font=\tiny] {\texttt{GET /article} + Bearer token} (8,-3.1);
  \draw[aarrow] (8.15,-3.7) -- (8.15,-4.1) -- (7.85,-4.1) -- (7.85,-3.7);
  \node[font=\tiny, text=black, right] at (8.3,-3.9) {Validate JWT via JWKS};
  \draw[adasharrow] (8,-4.5) -- node[alabel, above, font=\tiny, pos=0.4] {Content + \texttt{Aegon-Txn-Id} header} (0,-4.5);
  \draw[adotarrow] (8,-5.3) -- node[alabel, above, font=\tiny] {\texttt{POST /v1/content-hash}} (4,-5.3);

  \draw[adotarrow] (0,-6.2) -- node[alabel, above, font=\tiny] {\texttt{POST /v1/provenance}} (4,-6.2);
\end{tikzpicture}
\caption{License request flow showing the complete protocol sequence. All communication is standard HTTPS. Publisher validation requires only a JWKS endpoint---no broker roundtrip on content delivery.}
\label{fig:flow}
\end{figure*}

Publisher validation requires only a JWKS endpoint---no broker roundtrip on content delivery.

\subsection{Ledger and Audit Verification}

The broker maintains an append-only ledger (PostgreSQL with write-once rows). A Merkle tree is maintained over all transaction entries, following the Certificate Transparency model~\cite{rfc6962}:

\begin{itemize}
  \item \textbf{Signed Tree Head (STH):} The broker periodically publishes a signed Merkle root representing the current ledger state
  \item \textbf{Inclusion proofs:} For any transaction, the broker produces a Merkle inclusion proof showing the transaction is part of the committed tree
  \item \textbf{Third-party auditors} request the STH and verify inclusion proofs independently
\end{itemize}

\begin{figure*}[t]
\centering
\begin{tikzpicture}[node distance=1.0cm and 1.2cm]
  \node[anodebold, minimum width=4cm] (root) {Signed Tree Head (STH)};
  \node[anode, below left=1.2cm and 1.5cm of root, minimum width=2.8cm] (h12) {H(txn1 + txn2)};
  \node[anode, below right=1.2cm and 1.5cm of root, minimum width=2.8cm] (h34) {H(txn3 + txn4)};
  \node[anode, below left=1.2cm and 0.3cm of h12, minimum width=1.8cm] (t1) {\texttt{txn1}};
  \node[anode, below right=1.2cm and 0.3cm of h12, minimum width=1.8cm] (t2) {\texttt{txn2}};
  \node[anode, below left=1.2cm and 0.3cm of h34, minimum width=1.8cm] (t3) {\texttt{txn3}};
  \node[anode, below right=1.2cm and 0.3cm of h34, minimum width=1.8cm] (t4) {\texttt{txn4}};

  \draw[thick, aegonblue] (root) -- (h12);
  \draw[thick, aegonblue] (root) -- (h34);
  \draw[thick, aegonblue] (h12) -- (t1);
  \draw[thick, aegonblue] (h12) -- (t2);
  \draw[thick, aegonblue] (h34) -- (t3);
  \draw[thick, aegonblue] (h34) -- (t4);
\end{tikzpicture}
\caption{Merkle tree structure. The Signed Tree Head (STH) is the root hash. An auditor verifies an inclusion proof by recomputing hashes from any leaf to the root.}
\label{fig:merkle}
\end{figure*}
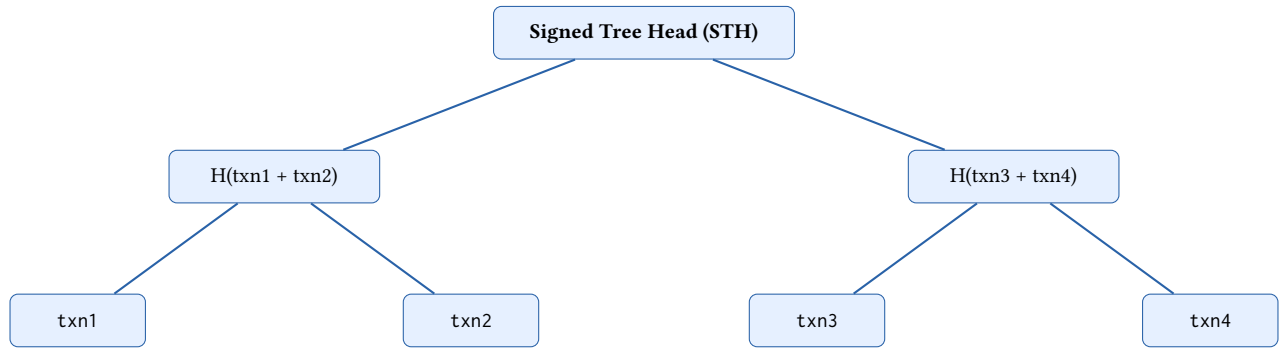

\textbf{Audit verification:} An auditor requests the current STH, then requests an inclusion proof for a specific \texttt{txn\_id}. The auditor verifies the Merkle path from the transaction leaf to the root. If valid, the transaction is confirmed as part of the committed ledger state.

\subsection{Broker Content Spot-Check}

The content hash recorded by the publisher is self-reported. To verify publishers are serving the content they claim, the broker independently spot-checks a configurable percentage of transactions (default: 5\%). For each spot-checked transaction, the broker independently fetches the content at the licensed \texttt{resource\_url}, computes a SHA-256 hash, and compares it against the publisher-reported hash and the platform-reported hash (from the \texttt{content\_fetched} provenance event). If all three hashes match, the transaction is verified. Persistent mismatches from a publisher trigger escalation. This operates on the same principle as tax audits: the broker does not verify every transaction, but the possibility of being spot-checked keeps publishers honest. At a 5\% spot-check rate, a publisher misreporting content hashes faces a 5\% per-transaction detection probability with consequences including audit escalation and platform suspension. For publishers whose primary goal is legitimate access revenue, the expected cost of detection and suspension plausibly exceeds the marginal gain from misreporting, providing a deterrence incentive analogous to tax audit compliance. A formal game-theoretic analysis is left as future work.

\textbf{Limitation:} Spot-checking works well for static content (news articles, blog posts). For highly dynamic content, the broker's independent fetch may return different content than what was served to the platform. The spot-check can be disabled per-publisher for dynamic content scopes.

\subsection{Provenance Event Log}

The platform SDK emits signed events at each AI transformation stage:

\begin{table}[htb]
\centering
\begin{tabularx}{\columnwidth}{l>{\raggedright\arraybackslash}X}
\toprule
\textbf{Event} & \textbf{When} \\
\midrule
\texttt{content\_fetched} & Content retrieved from publisher \\
\texttt{content\_chunked} & Content split into retrieval chunks \\
\texttt{chunk\_embedded} & Chunk converted to embedding \\
\texttt{chunk\_retrieved} & Chunk selected for RAG context \\
\texttt{content\_cited} & Content cited in final response \\
\bottomrule
\end{tabularx}
\caption{Provenance event types.}
\label{tab:provenance}
\end{table}

\begin{figure*}[t]
\centering
\begin{tikzpicture}[node distance=0.6cm]
  \node[anode, minimum width=2.2cm] (f) {Fetched};
  \node[anode, minimum width=2.2cm, right=0.6cm of f] (c) {Chunked};
  \node[anode, minimum width=2.2cm, right=0.6cm of c] (e) {Embedded};
  \node[anode, minimum width=2.2cm, right=0.6cm of e] (r) {Retrieved};
  \node[anode, minimum width=2.2cm, right=0.6cm of r] (ci) {Cited};
  \node[anode, minimum width=2.2cm, below=1.2cm of e] (led) {Ledger};

  \draw[aarrow] (f) -- (c);
  \draw[aarrow] (c) -- (e);
  \draw[aarrow] (e) -- (r);
  \draw[aarrow] (r) -- (ci);

  \draw[adotarrow] (f.south) -- (led);
  \draw[adotarrow] (c.south) -- (led);
  \draw[adotarrow] (e.south) -- (led);
  \draw[adotarrow] (r.south) -- (led);
  \draw[adotarrow] (ci.south) -- (led);
\end{tikzpicture}
\caption{Provenance chain. Each transformation stage emits a signed event (dotted) to the ledger, bound by \texttt{txn\_id}.}
\label{fig:provenance}
\end{figure*}
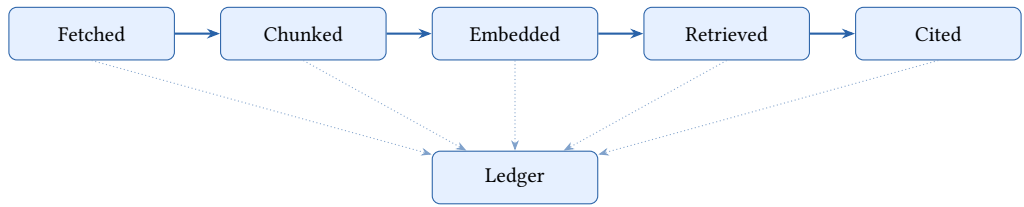

Each event carries the \texttt{txn\_id} from the original license, a content fingerprint at that stage, and the platform's signature.

\textbf{Honest limitation:} The provenance log is tamper-evident for recorded events. However, it cannot guarantee completeness---an adversarial platform can omit events. V1 enforcement is contractual.

\subsection{Replay Protection}
\label{sec:replay}

Each token includes a \texttt{jti} (JWT ID) and \texttt{exp} (expiration) claim. For \texttt{single\_use} licenses, publishers track seen \texttt{jti} values. To bound storage: (1)~short-lived tokens (5-minute TTL), enabling time-window deduplication; (2)~Bloom filters for \texttt{jti} deduplication within the TTL window. For \texttt{session} and \texttt{time\_bound\_cache} licenses, the \texttt{exp} claim alone is sufficient.

\section{Android Mobile Attestation}
\label{sec:android}

\subsection{Motivation}

On-device AI agents access web content during inference on Android devices. Unlike server-side platforms, mobile agents operate under intermittent connectivity, battery constraints, and no persistent server-side storage.

Android StrongBox~\cite{strongbox}---a discrete tamper-resistant secure element, architecturally distinct from the TEE---provides a hardware root of trust. Keys generated inside StrongBox cannot be exported, and the Key Attestation API~\cite{key-attestation} produces a certificate chain allowing any verifier to confirm hardware-backed key generation.

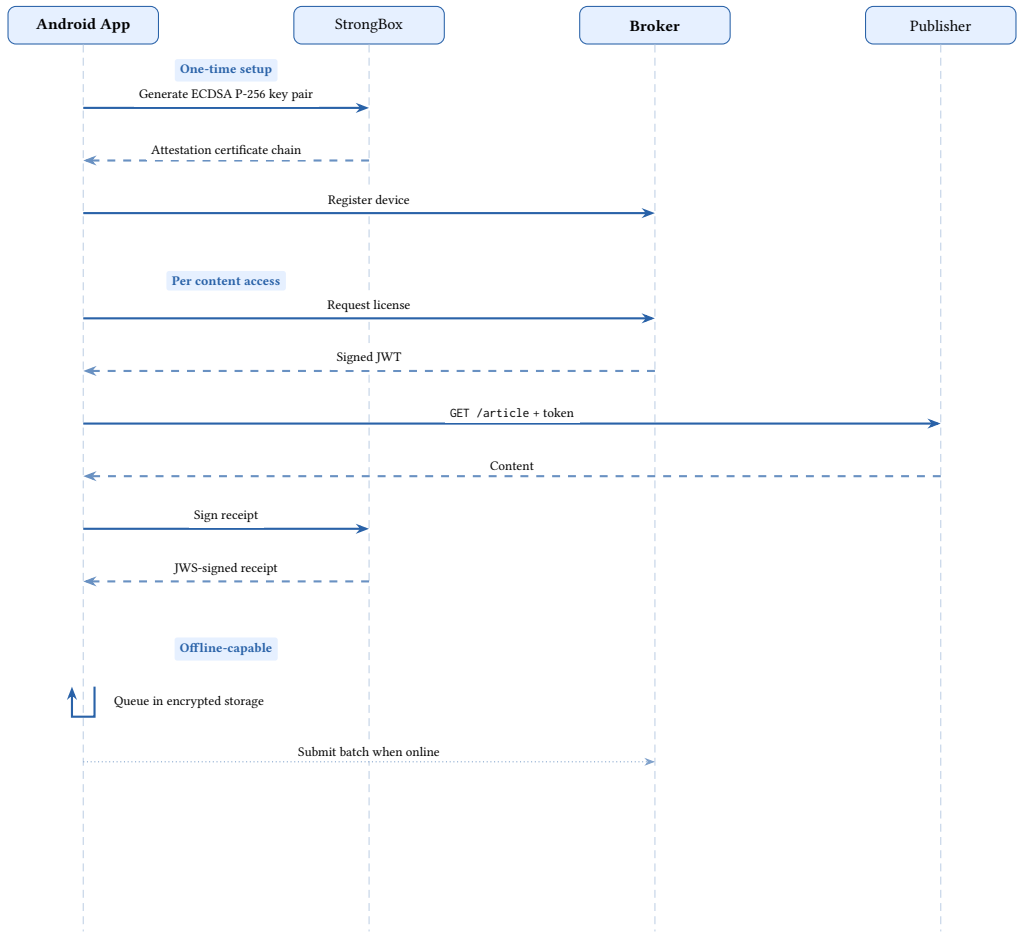
\begin{figure*}[t]
\centering
\begin{tikzpicture}
  \node[anodebold, minimum width=2cm, minimum height=0.5cm, font=\scriptsize\bfseries] (App) at (0,0) {Android App};
  \node[anode, minimum width=2cm, minimum height=0.5cm, font=\scriptsize] (SB) at (3.8,0) {StrongBox};
  \node[anodebold, minimum width=2cm, minimum height=0.5cm, font=\scriptsize\bfseries] (B) at (7.6,0) {Broker};
  \node[anode, minimum width=2cm, minimum height=0.5cm, font=\scriptsize] (Pub) at (11.4,0) {Publisher};

  \draw[densely dashed, aegonblue!30, thin] (App.south) -- ++(0,-11.8);
  \draw[densely dashed, aegonblue!30, thin] (SB.south) -- ++(0,-11.8);
  \draw[densely dashed, aegonblue!30, thin] (B.south) -- ++(0,-11.8);
  \draw[densely dashed, aegonblue!30, thin] (Pub.south) -- ++(0,-11.8);

  \node[font=\tiny\bfseries, text=aegonblue, fill=aegonlight, inner sep=2pt, rounded corners=1pt] at (1.9,-0.6) {One-time setup};
  \draw[aarrow] (0,-1.1) -- node[alabel, above, font=\tiny] {Generate ECDSA P-256 key pair} (3.8,-1.1);
  \draw[adasharrow] (3.8,-1.8) -- node[alabel, above, font=\tiny] {Attestation certificate chain} (0,-1.8);
  \draw[aarrow] (0,-2.5) -- node[alabel, above, font=\tiny] {Register device} (7.6,-2.5);

  \node[font=\tiny\bfseries, text=aegonblue, fill=aegonlight, inner sep=2pt, rounded corners=1pt] at (1.9,-3.4) {Per content access};
  \draw[aarrow] (0,-3.9) -- node[alabel, above, font=\tiny] {Request license} (7.6,-3.9);
  \draw[adasharrow] (7.6,-4.6) -- node[alabel, above, font=\tiny] {Signed JWT} (0,-4.6);
  \draw[aarrow] (0,-5.3) -- node[alabel, above, font=\tiny] {\texttt{GET /article} + token} (11.4,-5.3);
  \draw[adasharrow] (11.4,-6.0) -- node[alabel, above, font=\tiny] {Content} (0,-6.0);
  \draw[aarrow] (0,-6.7) -- node[alabel, above, font=\tiny] {Sign receipt} (3.8,-6.7);
  \draw[adasharrow] (3.8,-7.4) -- node[alabel, above, font=\tiny] {JWS-signed receipt} (0,-7.4);

  \node[font=\tiny\bfseries, text=aegonblue, fill=aegonlight, inner sep=2pt, rounded corners=1pt] at (1.9,-8.3) {Offline-capable};
  \draw[aarrow] (0.15,-8.8) -- (0.15,-9.2) -- (-0.15,-9.2) -- (-0.15,-8.8);
  \node[font=\tiny, text=black, right] at (0.3,-9.0) {Queue in encrypted storage};
  \draw[adotarrow] (0,-9.8) -- node[alabel, above, font=\tiny] {Submit batch when online} (7.6,-9.8);
\end{tikzpicture}
\caption{Android attestation flow: device setup (Phase~1), per-access receipt generation (Phase~2), and offline-capable batch submission (Phase~3). All participants shown with their interactions.}
\label{fig:android}
\end{figure*}

\subsection{Receipt Format}

The compliance receipt format is shown in Listing~\ref{lst:receipt}.

\begin{lstlisting}[language={},float=htb,caption={Hardware-attested compliance receipt format.},label={lst:receipt}]
{
  "receipt_id": "rcpt_xyz789",
  "txn_id": "txn_abc123",
  "publisher_scope_id": "ps_anon_12345",
  "timestamp": "2026-06-15T10:00:00Z",
  "event_type": "content_consumed",
  "content_hash": "sha256:abc123...",
  "license_constraints": {
    "license_type": "single_use",
    "training_allowed": false,
    "storage_policy": "ephemeral"
  },
  "device_attestation": {
    "key_id": "aegon_device_key",
    "strongbox_backed": true,
    "android_version": 14,
    "security_patch_level": "2026-02-01"
  }
}
\end{lstlisting}

Receipts are signed using JWS~\cite{rfc7515} with the device private key. \textbf{Privacy:} Receipts use a \texttt{publisher\_scope\_id}---a per-publisher pseudonymous identifier derived from the device key and publisher domain---preventing cross-publisher device tracking.

\subsection{Key Generation and Attestation}

Device keys are generated using \texttt{KeyPairGenerator} with \linebreak \mbox{\texttt{setIsStrongBoxBacked(true)}} and \mbox{\texttt{setAttestationChallenge}}, per the Android CDD~\cite{android-cdd}.

\textbf{StrongBox vs KeyMint}~\cite{keymint}: StrongBox runs in a discrete secure element with stronger isolation. KeyMint runs in the TEE. Both produce valid attestation chains, but StrongBox carries higher assurance (\texttt{SecurityLevel.STRONGBOX} vs \texttt{TRUSTED\_ENVIRONMENT}). The SDK uses StrongBox where available and falls back to KeyMint.

\textbf{Key revocation:} Google can revoke hardware attestation keys for compromised devices by updating its attestation key revocation list. If revoked, new receipts from that device fail attestation chain verification. Previously submitted and verified receipts remain valid (revocation is not retroactive). The device must generate new keys and re-register with the broker.

\subsection{Broker Verification}

\begin{enumerate}
  \item Verify attestation certificate chain to Google hardware attestation root CA
  \item Check \texttt{securityLevel} (StrongBox preferred, KeyMint accepted)
  \item Check \texttt{verifiedBootState}---reject unlocked bootloaders
  \item Verify JWS signature using device public key
  \item Check \texttt{txn\_id} exists in ledger; \texttt{content\_hash} matches publisher-reported hash
  \item Deduplicate on \texttt{receipt\_id}
\end{enumerate}

\subsection{Offline Proof Batching}

Receipts are persisted to encrypted local storage (SQLite with SQLCipher) and submitted in batches of up to 100. Retry uses exponential backoff (base 1s, max 60s, jitter). The broker rejects receipts with timestamps more than 7 days old.

\section{Reference Architecture}
\label{sec:impl}

The protocol is designed for implementation using standard, widely-available components.

\begin{table}[htb]
\centering
\begin{tabularx}{\columnwidth}{l>{\raggedright\arraybackslash}X}
\toprule
\textbf{Component} & \textbf{Technology} \\
\midrule
Broker & FastAPI (Python 3.11), deployable on any ASGI server \\
Ledger & Supabase PostgreSQL, write-once rows, JWKS rotation \\
Key management & AWS KMS (HSM-backed broker signing key) \\
Publisher middleware & Cloudflare Workers (edge validation, JWKS cached 5-min TTL) \\
Platform SDK & Python 3.11 + httpx, Node.js (TypeScript), MCP server integration \\
Android client & Kotlin, API 31+, StrongBox + KeyMint fallback, SQLCipher offline queue \\
\bottomrule
\end{tabularx}
\caption{Proposed implementation components.}
\label{tab:impl}
\end{table}

\textbf{Adoption model.} The design prioritizes near-zero integration friction. The MCP server~\cite{mcp} exposes license acquisition, provenance reporting, and receipt submission as standard MCP tool calls---any MCP-compatible AI agent (e.g., Claude, GPT-based agents) can acquire Aegon licenses without protocol-specific integration beyond standard MCP tool wiring. For publishers, the Cloudflare Workers middleware is a single copy-paste script: it fetches and caches the broker's JWKS endpoint, validates incoming JWT tokens at the CDN edge, and injects \texttt{Aegon-Txn-Id} response headers---no SDK installation, no server-side changes. This zero-friction model is deliberate: OAuth~2.0~\cite{rfc6749} succeeded in part because client libraries existed for every language and framework; Aegon targets the same adoption dynamic for content licensing.

\section{Evaluation Plan}
\label{sec:eval}

The following experiments are planned against a prototype implementation. The evaluation methodology and performance targets described here define what ``good'' looks like for this class of protocol and establish the baseline against which future implementations can be compared. Measured results will be published in a follow-on paper.

\begin{table}[htb]
\centering
\begin{tabularx}{\columnwidth}{l>{\raggedright\arraybackslash}X}
\toprule
\textbf{Experiment} & \textbf{Methodology} \\
\midrule
Token issuance latency & P50/P95/P99 at 1, 10, 100, 1000 req/s \\
Token validation latency & Cached vs uncached JWKS, edge vs origin \\
Provenance overhead & Baseline fetch vs fetch + provenance tracking \\
Ledger throughput & Sustained write rate, P95 write latency \\
Attack detection & Token replay, forgery, unauthorized reuse \\
Cost per transaction & Infrastructure cost at 10K / 1M / 100M txn/day \\
\midrule
StrongBox signing & P50/P95/P99 on Pixel 8 (StrongBox) vs Pixel 6 (KeyMint-only) \\
Receipt size & Total bytes: JWS envelope + attestation chain + metadata \\
Offline queue & Queue depth vs submission latency, retry with backoff \\
Battery impact & Energy per receipt via Android Battery Historian \\
Attestation verification & Server-side certificate chain verification latency \\
\bottomrule
\end{tabularx}
\caption{Web layer (top) and Android layer (bottom) experiments.}
\label{tab:experiments}
\end{table}

\begin{table}[htb]
\centering
\begin{tabularx}{\columnwidth}{l>{\raggedright\arraybackslash}X}
\toprule
\textbf{Metric} & \textbf{Target} \\
\midrule
Token validation (cached JWKS, edge) & $<$ 10ms P95 \\
Token issuance (P95, 100 req/s) & $<$ 50ms \\
Provenance event overhead & $<$ 5ms per event \\
StrongBox signing & $<$ 100ms P95~\cite{tee-survey} \\
KeyMint signing & $<$ 20ms P95 \\
Receipt size & $<$ 4KB \\
Forgery/replay vectors & All detected \\
\bottomrule
\end{tabularx}
\caption{Expected performance targets.}
\label{tab:targets}
\end{table}

\section{Discussion}
\label{sec:discussion}

\subsection{Limitations}

\textbf{Provenance completeness.} Tamper-evident for recorded events but cannot guarantee completeness against an adversarial platform.

\textbf{No-training enforcement.} Proving content was not used for training is an open problem. The protocol provides contractual enforcement.

\textbf{Android coverage.} StrongBox availability varies by manufacturer. Devices without StrongBox fall back to KeyMint (lower assurance).

\textbf{App integrity.} Key Attestation proves key hardware-binding but not app integrity. Combining with Play Integrity API~\cite{play-integrity} is planned.

\textbf{Broker trust.} The broker is a trusted third party (Section~\ref{sec:threat}). HSM-backed keys and audit access mitigate but do not eliminate this dependency.

\subsection{Relationship to Existing Standards}

Aegon complements existing standards:
\begin{itemize}
  \item \textbf{RSL}~\cite{rsl} declares licensing policies. Aegon adds the audit trail.
  \item \textbf{Peek-Then-Pay}~\cite{peek-then-pay} optimizes discovery and pricing. Aegon provides post-transaction verification.
  \item \textbf{PEAC}~\cite{peac} generates compliance receipts. Aegon's Merkle ledger provides independently verifiable audit proofs.
  \item \textbf{OpenAttribution}~\cite{openattribution} tracks content telemetry. Aegon binds provenance to licensing transactions.
\end{itemize}

Standardization path: register \texttt{Aegon-License}, \texttt{Aegon-Txn-Id}, and \texttt{Aegon-Provenance} HTTP headers via IETF/IANA; submit Internet-Draft for the token claim format building on Agentic JWT~\cite{agentic-jwt}; engage W3C on provenance vocabulary aligned with C2PA~\cite{c2pa} and OpenAttribution~\cite{openattribution}.

\subsection{Future Work}
\label{sec:future}

\begin{itemize}
  \item \textbf{RSL integration:} Map RSL license types to Aegon token claims
  \item \textbf{iOS extension:} Apple App Attest~\cite{app-attest} and Secure Enclave
  \item \textbf{Information Isotope integration}~\cite{info-isotopes}: Combined ex-ante licensing with post-hoc detection
  \item \textbf{ZK proofs:} Pedersen commitments for private compliance audits
  \item \textbf{TEE-wrapped inference:} Attest that inference ran under a content policy
  \item \textbf{CDN-native validation:} Building on Cloudflare~\cite{cloudflare-paypercrawl}
  \item \textbf{Distributed broker:} Eliminate single trust assumption via MPC
\end{itemize}

\section{Conclusion}
\label{sec:conclusion}

We have presented Aegon, a protocol design for auditable AI content licensing records. Aegon defines two distinct layers. The web layer specifies content-specific JWT licensing claim semantics and a Certificate Transparency-style Merkle ledger enabling independent audit verification via inclusion proofs, with a signed provenance event log tracking content through AI transformation stages. The Android layer introduces hardware-attested compliance receipts for on-device AI agents, using Android StrongBox secure element attestation with offline-capable batch submission, replay resistance, and per-publisher pseudonymous identifiers---to our knowledge, the first application of hardware attestation to AI content licensing on mobile devices.

The demand for auditable AI content licensing infrastructure is validated by ongoing litigation and the rapid emergence of commercial platforms in this space. Aegon addresses a specific gap: tamper-evident transaction records and independently verifiable audit proofs. Provenance completeness and training misuse prevention remain open problems, and enforcement in v1 is partially contractual. A prototype implementation and full performance evaluation are planned as future work.


\bibliographystyle{ACM-Reference-Format}
\bibliography{references}

@misc{nyt-v-openai,
  title   = {The New York Times Company v. Microsoft Corporation et al.},
  note    = {Case No. 1:23-cv-11195 (S.D.N.Y. 2023)},
  year    = {2023},
  key     = {NYT},
}

@misc{getty-v-stability,
  title   = {Getty Images (US), Inc. v. Stability AI, Inc.},
  note    = {Case No. 1:23-cv-00135 (D. Del. 2023)},
  year    = {2023},
  key     = {Getty},
}

@misc{authorsguild-v-openai,
  title   = {Authors Guild et al. v. OpenAI Inc. et al.},
  note    = {Case No. 1:23-cv-08292 (S.D.N.Y. 2023)},
  year    = {2023},
  key     = {AuthorsGuild},
}

@misc{newscorp-openai,
  author  = {Benjamin Mullin and Keach Hagey},
  title   = {News Corp Signs AI Deal With OpenAI Worth More Than \$250 Million},
  howpublished = {Wall Street Journal},
  year    = {2024},
  month   = {May},
}

@misc{gemini-nano,
  author  = {Google},
  title   = {Gemini Nano: On-Device AI},
  howpublished = {Android Developers Blog},
  year    = {2024},
}

@misc{llama3,
  author  = {Meta},
  title   = {Llama 3: Open Foundation and Fine-Tuned Models},
  year    = {2024},
}

@misc{rfc7517,
  author  = {Michael Jones},
  title   = {{JSON Web Key (JWK)}},
  howpublished = {RFC 7517, IETF},
  year    = {2015},
}

@misc{rfc7519,
  author  = {Michael Jones and John Bradley and Nat Sakimura},
  title   = {{JSON Web Token (JWT)}},
  howpublished = {RFC 7519, IETF},
  year    = {2015},
}

@misc{rfc6962,
  author  = {Ben Laurie and Adam Langley and Emilia Kasper},
  title   = {{Certificate Transparency}},
  howpublished = {RFC 6962, IETF},
  year    = {2013},
}

@misc{strongbox,
  author  = {{Android Developers}},
  title   = {Hardware-backed Keystore: StrongBox},
  howpublished = {developer.android.com},
  year    = {2024},
}

@misc{key-attestation,
  author  = {{Android Developers}},
  title   = {Key Attestation},
  howpublished = {developer.android.com},
  year    = {2024},
}

@misc{widevine,
  author  = {Google},
  title   = {Widevine DRM Architecture Overview},
  howpublished = {widevine.com},
  year    = {2024},
}

@misc{rfc6749,
  author  = {Dick Hardt},
  title   = {{The OAuth 2.0 Authorization Framework}},
  howpublished = {RFC 6749, IETF},
  year    = {2012},
}

@misc{rfc9309,
  author  = {Martijn Koster and Gary Illyes and Henner Zeller and Lizzi Sassman},
  title   = {{Robots Exclusion Protocol}},
  howpublished = {RFC 9309, IETF},
  year    = {2022},
}

@misc{oidc,
  author  = {Nat Sakimura and others},
  title   = {{OpenID Connect Core 1.0}},
  howpublished = {OpenID Foundation},
  year    = {2014},
}

@misc{c2pa,
  author  = {{C2PA}},
  title   = {{C2PA Technical Specification v2.2}},
  howpublished = {c2pa.org},
  year    = {2025},
  month   = {May},
}

@misc{blockchain-drm,
  author  = {J. Xu and others},
  title   = {Blockchain-Based Digital Rights Management: A Survey},
  howpublished = {IEEE Access},
  year    = {2021},
}

@article{timestamping,
  author  = {Stuart Haber and W. Scott Stornetta},
  title   = {How to Time-Stamp a Digital Document},
  journal = {Journal of Cryptology},
  volume  = {3},
  number  = {2},
  pages   = {99--111},
  year    = {1991},
}

@misc{keymint,
  author  = {{Android Developers}},
  title   = {KeyMint HAL},
  howpublished = {source.android.com},
  year    = {2024},
}

@misc{play-integrity,
  author  = {Google},
  title   = {Play Integrity API},
  howpublished = {developer.android.com},
  year    = {2024},
}

@misc{app-attest,
  author  = {Apple},
  title   = {App Attest},
  howpublished = {developer.apple.com},
  year    = {2024},
}

@misc{rfc7515,
  author  = {Michael Jones and John Bradley and Nat Sakimura},
  title   = {{JSON Web Signature (JWS)}},
  howpublished = {RFC 7515, IETF},
  year    = {2015},
}

@misc{android-cdd,
  author  = {{Android Compatibility Definition Document}},
  title   = {Section 9.11: Keys and Credentials},
  howpublished = {source.android.com},
  year    = {2024},
}

@misc{tee-survey,
  author  = {Mohamed Sabt and Mohammed Achemlal and Abdelmadjid Bouabdallah},
  title   = {Trusted Execution Environments: A Survey},
  howpublished = {IEEE Communications Surveys and Tutorials},
  year    = {2015},
}

@misc{rsl,
  author  = {Eckart Walther and R.V. Guha},
  title   = {{RSL 1.0: Really Simple Licensing}},
  howpublished = {RSL Collective, rslstandard.org},
  year    = {2025},
  month   = {December},
}

@misc{aibdp,
  author  = {A. Jewell},
  title   = {{AI Boundary Declaration Protocol (AIBDP)}},
  howpublished = {IETF Internet-Draft draft-jewell-aibdp-00},
  year    = {2025},
  month   = {August},
}

@misc{agentic-jwt,
  author  = {S. Goswami and others},
  title   = {{Agentic JWT}},
  howpublished = {IETF Internet-Draft draft-goswami-agentic-jwt-00},
  year    = {2025},
  month   = {December},
  note    = {arXiv:2509.13597},
}

@misc{peek-then-pay,
  author  = {{FetchRight}},
  title   = {{Peek-Then-Pay: HTTP 203-Based Content Licensing Protocol}},
  howpublished = {fetchright.ai},
  year    = {2025},
  note    = {Accessed March 2026. No formal IETF draft or arXiv specification; citation refers to working implementation documentation},
}

@misc{peac,
  author  = {{PEAC Protocol}},
  title   = {{Verifiable Interaction Records for AI Content Compliance}},
  howpublished = {peacprotocol.org},
  year    = {2025},
  note    = {Accessed March 2026. No formal IETF draft or arXiv specification; citation refers to working implementation documentation},
}

@misc{openattribution,
  author  = {{OpenAttribution}},
  title   = {{AIMS: AI Manifest Standard and Telemetry}},
  howpublished = {openattribution.dev},
  year    = {2025},
}

@misc{cloudflare-paypercrawl,
  author  = {{Cloudflare}},
  title   = {{Pay-Per-Crawl: AI Bot Content Monetization}},
  howpublished = {blog.cloudflare.com},
  year    = {2025},
  month   = {July},
}

@misc{tollbit,
  author  = {{TollBit}},
  title   = {{AI Content Licensing Platform}},
  howpublished = {tollbit.com},
  year    = {2024},
}

@misc{cashmere,
  author  = {{Cashmere}},
  title   = {{AI Content Management for Premium Publishers}},
  howpublished = {cashmere.io},
  year    = {2026},
}

@misc{veritrail,
  author  = {{Microsoft Research}},
  title   = {{VeriTrail: DAG-Based Provenance Tracing for LLM Hallucination Detection}},
  howpublished = {ICLR 2026},
  year    = {2026},
}

@misc{fg-trac,
  title   = {{FG-Trac: Fine-Grained Traceability for ML Pipelines with Cryptographic Commitments}},
  howpublished = {arXiv},
  year    = {2025},
  key     = {FG-Trac},
}

@misc{info-isotopes,
  author  = {{Qi et al.}},
  title   = {{Information Isotopes: Auditing AI Training Data with Cryptographic Markers}},
  howpublished = {Nature Communications},
  year    = {2026},
  month   = {February},
}

@misc{ibis,
  author  = {{CSIRO Data61 and others}},
  title   = {{IBIS: Blockchain Framework for Ethical AI Model Training and Copyright Compliance}},
  howpublished = {arXiv:2404.06077},
  year    = {2024},
  month   = {April},
}

@misc{content-arcs,
  author  = {{University of Surrey DECaDE Centre}},
  title   = {{Content ARCs: Decentralized Content Rights in the Age of Generative AI}},
  howpublished = {arXiv:2503.14519},
  year    = {2025},
  month   = {March},
}

@misc{mcp,
  author  = {{Anthropic}},
  title   = {{Model Context Protocol (MCP) Specification}},
  howpublished = {modelcontextprotocol.io},
  year    = {2025},
}

\end{document}